\documentclass[twocolumn,english,showpacs]{revtex4}\newcommand{\JHEPonly}[1]{}
\usepackage{graphicx}
\usepackage{amsmath,amssymb,bm}

\def\be{\begin{equation}}
\def\ee{\end{equation}}
\def\bea{\begin{eqnarray}}
\def\eea{\end{eqnarray}}

\def\re#1{Re(#1)}
\def\im#1{Im(#1)}
\def\K{{\cal K}}
\def\Order#1{{\cal O}\left(#1\right)}

\begin{document}

\title{Correspondence between quasinormal modes and grey-body factors for massive fields in Schwarzschild-de Sitter spacetime}
\author{Zainab Malik}\email{zainabmalik8115@outlook.com}
\affiliation{Institute of Applied Sciences and Intelligent Systems, H-15, Pakistan}

\begin{abstract}
Recently, a correspondence between quasinormal modes and grey-body factors of black holes has been established. This correspondence is known to be exact in the eikonal regime for a large class of asymptotically flat black holes and approximate when the multipole number \( \ell \) is small. In this work, we demonstrate that there exists a regime where the correspondence holds with unprecedented accuracy even for the lowest multipole numbers: specifically, for perturbations of massive fields in the background of asymptotically de Sitter black holes, provided the field mass is not very small. 
We also fill the gap in the existing literature via finding the grey-body factors of a massive scalar field in the Schwarzschild- de Sitter background, when $\mu M/m_{P}$ is not small.
\end{abstract}

\pacs{04.30.Nk,04.50.+h}
\maketitle

\section{Introduction}

The study of black hole perturbations plays a crucial role in understanding the fundamental properties of black holes and their interactions with surrounding matter and fields. Quasinormal modes (QNMs) \cite{Kokkotas:1999bd,Nollert:1999ji,Konoplya:2011qq}, which describe the damped oscillations of black holes under perturbations, have become a central topic in modern theoretical physics. These modes are not only essential for the characterization of black holes in astrophysical contexts, such as gravitational wave observations \cite{LIGOScientific:2016aoc,LIGOScientific:2017vwq,LIGOScientific:2020zkf,Babak:2017tow}, but also provide insights into the stability of spacetime geometries and their underlying physical properties.

On the other hand, grey-body factors describe the partial transmission of radiation through the gravitational potential barrier surrounding a black hole \cite{Page:1976df,Page:1976ki,Kanti:2004nr}. These factors quantify the deviation from perfect blackbody radiation due to the scattering of emitted particles by the black hole's effective potential. Understanding grey-body factors is critical for studying black hole evaporation via Hawking radiation and provides important constraints for models of quantum gravity and high-energy astrophysics.

Recently, a remarkable correspondence between quasinormal modes and grey-body factors has been established  and discussed for a wide class of black holes and wormholes \cite{Konoplya:2024lir,Oshita:2023cjz,Rosato:2024arw,Oshita:2024fzf,Konoplya:2024vuj,Dubinsky:2024vbn,Bolokhov:2024otn,Skvortsova:2024msa}. This correspondence reveals a deeper connection between the resonant frequencies of black holes and their scattering properties, highlighting the interplay between dynamical and radiative processes in curved spacetime. For many asymptotically flat black holes, this correspondence has been shown to be exact in the high multipole number ($\ell \rightarrow \infty$) limit \cite{Konoplya:2024lir}, where the WKB approximation for the effective potential becomes highly accurate. However, for small values of $\ell$, the correspondence holds only approximately, with deviations dependent on the specific spacetime geometry and field parameters \cite{Konoplya:2024lir,Konoplya:2024vuj,Skvortsova:2024msa}.

Quasinormal modes of various asymptotically de Sitter black holes have been studied in a great number of publications \cite{Zhidenko:2003wq,Konoplya:2004uk,Cuyubamba:2016cug,Dyatlov:2010hq,Jansen:2017oag,Molina:2003ff,Jing:2003wq,Aragon:2020qdc,Konoplya:2007zx,Mo:2018nnu,Konoplya:2013sba}, while fewer works were devoted to grey-body factors \cite{Pappas:2016ovo,Kanti:2017ubd,Pappas:2017kam}. To the best of our knowledge, no calculations of grey-body factors were done in the regime of large mass of the field: 
\begin{equation}
\frac{\mu M}{m_P} \gg 1.
\end{equation}
At the same time, this regime includes radiation of almost all massive particles of the Standard Model starting from the mass of the black hole of about $10^{22}$ kg, which corresponds to a tiny balck hole of $\sim 10^{-5}$ m.  

The study of spectral charactecterisitcs, such as quasinormal modes or grey-body factors of massive fields in the black hole background has its own motivation.  
When fields acquire an effective mass—whether intrinsically or due to external factors such as magnetic fields \cite{Konoplya:2007yy,Konoplya:2008hj,Wu:2015fwa,Davlataliev:2024mjl,Kokkotas:2010zd} or brane-world scenarios \cite{Seahra:2004fg} — the dynamics become even richer, leading to phenomena such as long-lived modes (quasibound states) 
\cite{Ohashi:2004wr,Zinhailo:2018ska,Fernandes:2021qvr,Konoplya:2005hr,
Konoplya:2017tvu,Churilova:2019qph,Bolokhov:2023bwm,Percival:2020skc,Zhidenko:2006rs}
and potential non-trivial change in change in the echoes \cite{Konoplya:2024wds}. These effective masses introduce new scales into the system, modifying the behavior of perturbations and offering a pathway to probe beyond-standard-model physics. Additionally, massive fields play a key role in exploring the interaction between black holes and dark matter candidates, as well as in understanding potential deviations from classical general relativity in the presence of new physics \cite{Cardoso:2019rvt}. In addition, massive fields must contribute into the very long wavelenegths observed via Pulsar Timing Array observations \cite{NANOGrav:2023hvm,NANOGrav:2023gor}, because of peculiar oscillatory and slowly decaying asymptotic tails \cite{Koyama:2001ee, Gibbons:2008gg, Konoplya:2006gq, Rogatko:2007zz, Jing:2004zb} appropriate to massive fields \cite{Konoplya:2023fmh}. 

In this work, we extend the analysis of the quasinormal mode and grey-body factor correspondence to black holes in asymptotically de Sitter spacetimes, considering perturbations of massive scalar fields. We show that in the regime where the mass of the field is not very small, the correspondence holds with unprecedented accuracy. This result provides a new perspective on the connection between quasinormal modes and grey-body factors and offers a deeper understanding of the scattering properties of black holes in cosmological settings.

The remainder of this paper is organized as follows. In Section II, we derive the quasinormal modes (QNMs) of a massive scalar field in the Schwarzschild-de Sitter background by applying the higher-order WKB method. Section III explores the computation of grey-body factors and their relation to the transmission coefficients across the black hole’s potential barrier. In Section IV, we investigate the correspondence between quasinormal modes and grey-body factors, emphasizing the role of scalar field mass and cosmological constant in enhancing the precision of this relationship. The extension of the correspondence to other asymptotically de Sitter spacetimes, such as Reissner-Nordström-de Sitter and Mannheim-Kazanas spacetimes is addressed in Section V. Finally, we summarize our findings and discuss their implications in Section VI.

\section{Quasinormal modes}

The Schwarzschild-de Sitter black hole metric is described by the following line element:
\begin{equation}
ds^2 = -f(r) dt^2 + \frac{dr^2}{f(r)} + r^2 \left( d\theta^2 + \sin^2\theta \, d\phi^2 \right), 
\end{equation}
where the metric function is given by:
\begin{equation}\label{MKsolution}
f(r) = 1 - \frac{2 M}{r}  - \frac{\Lambda r^2}{3}.
\end{equation}

The function $f(r)$ vanishes at the event horizon $r = r_{h}$ and at the de Sitter horizon $r = r_{dS}$.  

\begin{figure}
\resizebox{\linewidth}{!}{\includegraphics{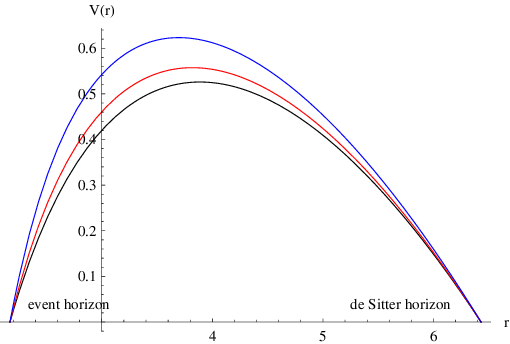}}
\caption{Effective potentials of a massive scalar field in the Schwarzschild-de Sitter spacetime: $\ell=0$, $1$, $2$ (from bottom to top), $\mu =1.5$, $M=1$, $\Lambda =0.05$.}\label{fig:Potentials}
\end{figure}

The general covariant form of the perturbation equation for a test scalar field is:
\begin{subequations}\label{coveqs}
\begin{eqnarray}\label{KGg}
\frac{1}{\sqrt{-g}}\partial_\mu \left(\sqrt{-g}g^{\mu \nu}\partial_\nu\Phi\right) &=& \mu^2 \Phi,
\end{eqnarray}
\end{subequations}
where $\mu$ denotes the mass of the scalar field.

Upon separation of variables, the perturbation equations for test scalar and electromagnetic fields reduce to a Schrödinger-like wave equation:
\begin{equation}\label{sp}
\frac{d^2\Psi}{dr_*^2} + \omega^2 \Psi - V(r) \Psi = 0,
\end{equation}
where the tortoise coordinate $r_*$ is defined as:
\begin{equation}
dr_* = \frac{dr}{f(r)}.
\end{equation}

The effective potential $V(r)$ is expressed as:
\begin{equation}
V(r) = \frac{f(r)}{r^2} \ell (\ell+1) +  \frac{1}{2r} \frac{d}{dr} \left( f(r)^2 \right) + \mu^2 f(r).
\end{equation}

This potential forms a barrier with a single peak, as illustrated in Fig. \ref{fig:Potentials}.

The boundary conditions for quasinormal modes require purely outgoing waves at the de Sitter horizon and purely ingoing waves at the event horizon (see, for example, \cite{Zhidenko:2003wq,Konoplya:2004uk}). Explicitly, this can be written as:
\begin{equation}
\Psi \sim e^{- i \omega r_*}, \quad r_* \rightarrow - \infty,
\end{equation}
\begin{equation}
\Psi \sim e^{+ i \omega r_*}, \quad r_* \rightarrow r_{dS}.
\end{equation}

The WKB method, frequently employed to compute quasinormal modes of black holes \cite{Schutz:1985km, Iyer:1986np, Konoplya:2003ii, Matyjasek:2017psv}, provides reliable results when the potential exhibits two turning points and resembles the form shown in Fig. \ref{fig:Potentials}. The quasinormal frequencies $\omega$ are determined by the condition:
\begin{equation}\label{WKB}
\frac{i Q_{0}}{\sqrt{2 Q_{0}''}} - \sum_{i=2}^{p} \Lambda_{i} - n - \frac{1}{2}=0, 
\end{equation}
where $\Lambda_i$ are correction terms \cite{Schutz:1985km, Iyer:1986np, Konoplya:2003ii, Matyjasek:2017psv} that depend on derivatives of the potential at its maximum. Here, $Q = \omega^2 - V$, and $Q_0^{(i)}$ represents the $i$-th derivative at the peak of the potential. The integer $n$ denotes the overtone number.

The accuracy of the WKB approach can be significantly enhanced by applying Padé approximants, which we also utilize to compute quasinormal modes with high precision. As demonstrated in \cite{Fontana:2020syy,Dubinsky:2024hmn}, while the WKB method is typically unreliable for massive fields, it achieves remarkable accuracy when both the mass term and the positive cosmological constant are present.

The WKB method has been extensively applied to compute quasinormal modes in numerous studies \cite{Zinhailo:2019rwd, Dubinsky:2024fvi, Konoplya:2005sy, Skvortsova:2024wly, Guo:2023ivz, Bolokhov:2023dxq, Paul:2023eep, Cuyubamba:2016cug, Gong:2023ghh, Zhidenko:2008fp, Zinhailo:2018ska, Skvortsova:2024atk, Malik:2024tuf, Malik:2024bmp, Bolokhov:2023bwm, Dubinsky:2024hmn,  Konoplya:2001ji, Dubinsky:2024rvf, Konoplya:2010kv}, consistently demonstrating excellent agreement with alternative techniques within their respective domains of validity.

The quasinormal modes of asymptotically de Sitter spacetimes differ qualitatively from those in asymptotically flat cases. In the latter, quasinormal modes do not form a complete set and are eventually overtaken by power-law tails. Conversely, in asymptotically de Sitter spacetimes, quasinormal modes govern the decay of perturbations even at late times \cite{Dyatlov:2010hq,Dubinsky:2024jqi}.

\section{Grey-body factors}

\begin{figure*}
\resizebox{\linewidth}{!}{\includegraphics{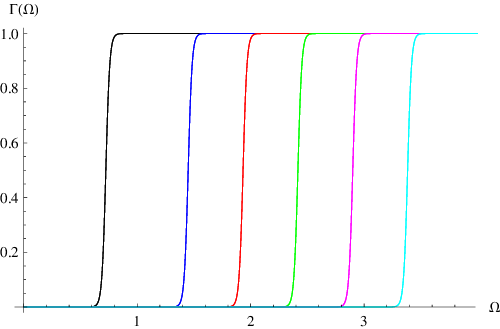}~~\includegraphics{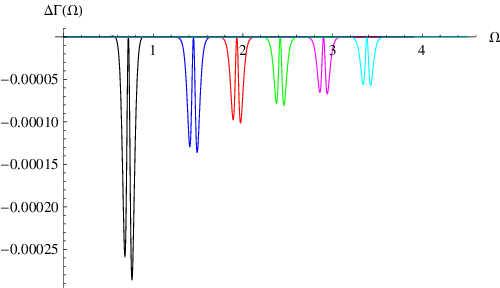}}
\caption{Grey-body factors obtained by the higher WKB method and via the correspondence with quasinormal modes (left) and relative difference between them (right) for $\ell = 0$, $\mu = 1.5$, $3$, $4$, $5$, $6$, $7$ (from left to right), $M = 1$, $\Lambda = 0.05$.}\label{fig:Lambda005L0}
\end{figure*}

\begin{figure*}
\resizebox{\linewidth}{!}{\includegraphics{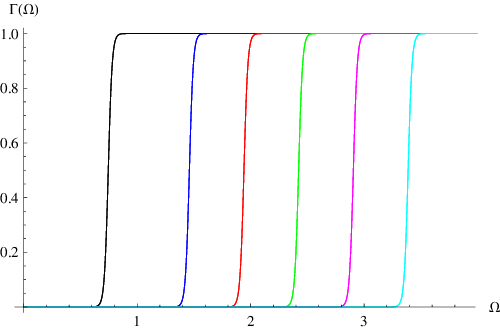}~~\includegraphics{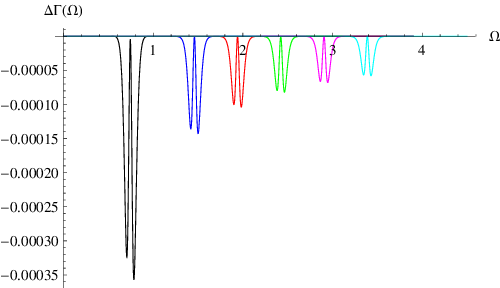}}
\caption{Grey-body factors obtained by the higher WKB method and via the correspondence with quasinormal modes (left) and relative difference (right) for $\ell = 1$, $\mu = 1.5$, $3$, $4$, $5$, $6$, $7$ (from left to right), $M = 1$, $\Lambda = 0.05$.}\label{fig:Lambda005L1}
\end{figure*}

\begin{figure*}
\resizebox{\linewidth}{!}{\includegraphics{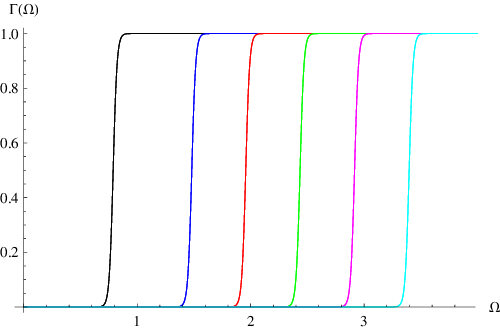}~~\includegraphics{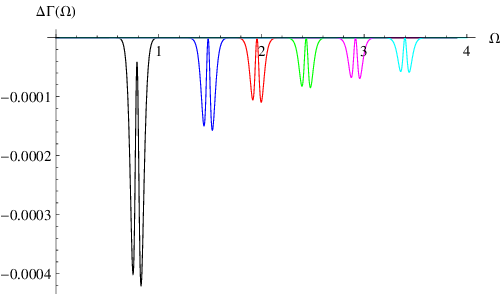}}
\caption{Grey-body factors obtained by the higher WKB method and via the correspondence with quasinormal modes (left) and relative difference between them (right) for $\ell = 2$, $\mu = 1.5$, $3$, $4$, $5$, $6$, $7$ (from left to right), $M = 1$, $\Lambda = 0.05$.}\label{fig:Lambda005L2}
\end{figure*}

\begin{figure*}
\resizebox{\linewidth}{!}{\includegraphics{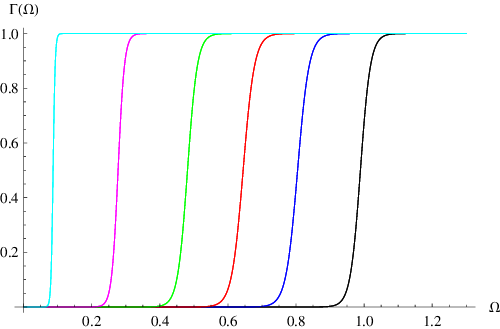}~~\includegraphics{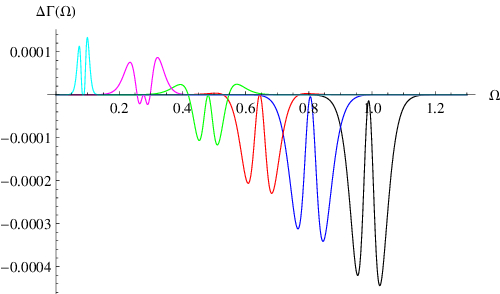}}
\caption{Grey-body factors obtained by the higher WKB method and via the correspondence with quasinormal modes (left) and relative difference between them (right) for $\ell = 0$, $\mu = 1.5$, $\Lambda =0.2$, $0.4$, $0.6$, $0.8$, $1.0$, $1.1$ (from right to left), $M = 1$, $\Lambda = 0.05$.}\label{fig:mu1point5}
\end{figure*}

\begin{figure*}
\resizebox{\linewidth}{!}{\includegraphics{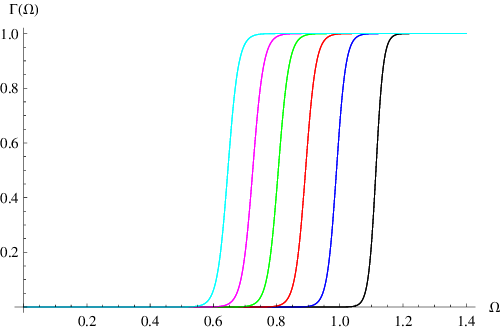}~~\includegraphics{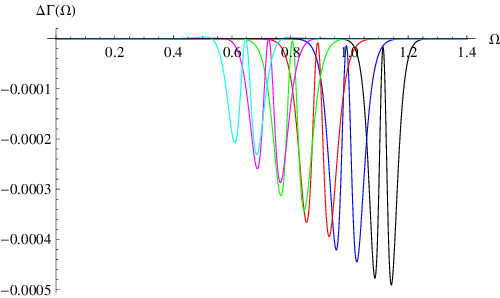}}
\caption{Grey-body factors of RNdS black hole obtained by the higher WKB method and via the correspondence with quasinormal modes (left) and relative difference between them (right) for $\ell = 0$, $\mu = 1.5$, $Q=0.1$, $\Lambda =0.1$, $0.2$, $0.3$, $0.4$, $0.5$, $0.6$ (from right to left), $M = 1$, $\Lambda = 0.05$.}\label{fig:mu1point5Charged}
\end{figure*}

Grey-body factors are intimately connected to the transmission coefficients that quantify the fraction of radiation capable of surmounting the potential barrier and reaching a distant observer. Due to the inherent symmetry of the scattering problem, the boundary conditions governing this process are expressed as:
\begin{equation}
\begin{array}{rclcl}
\Psi &=& e^{-i\Omega r_*} + R e^{i\Omega r_*}, &\quad& r_* \to +\infty, \\
\Psi &=& T e^{-i\Omega r_*}, &\quad& r_* \to -\infty,
\end{array}
\end{equation}
where $R$ and $T$ represent the reflection and transmission coefficients, respectively. In accordance with recent studies \cite{Konoplya:2024lir}, we differentiate between the continuous, purely real frequency $\Omega$ that characterizes scattering phenomena and the discrete, complex quasinormal modes $\omega_n$, which correspond to the intrinsic resonances of the black hole.

The grey-body factor, indicative of the transmission efficiency, is formally defined by:
\begin{equation}
\Gamma_{\ell}(\Omega) = |T|^2 = 1 - |R|^2.
\end{equation}

The WKB approximation adopts the following expression for the grey-body factor:
\begin{equation}\label{eq:gbfactor}
\Gamma_{\ell}(\Omega) = \frac{1}{1 + e^{2\pi i \K}}.
\end{equation}
This formulation has been extensively employed to derive grey-body factors in numerous contemporary investigations \cite{Dubinsky:2024nzo, Dubinsky:2024vbn, Toshmatov:2015wga, Skvortsova:2024msa, Konoplya:2020jgt}.

The relationship between grey-body factors $\Gamma_{\ell}(\Omega)$ and the fundamental quasinormal mode $\omega_0$ can be expressed through the transmission coefficient as:
\begin{equation}\label{transmission-eikonal}
\Gamma_{\ell}(\Omega) \equiv |T|^2 = \left(1 + e^{2\pi \frac{\Omega^2 - \re{\omega_0}^2}{4 \re{\omega_0} \im{\omega_0}}}\right)^{-1} + \Order{\ell^{-1}}.
\end{equation}
This relation is exact in the eikonal limit $\ell \to \infty$ and serves as a reliable approximation even for small values of $\ell$. Our analysis reveals that this connection remains valid beyond the eikonal regime with high accuracy when the mass of the field and cosmological constant are turned on.

Incorporating corrections up to the third order, we obtain the following refined expression \cite{Konoplya:2024lir}:
\begin{eqnarray}\nonumber
&&i \K=\frac{\Omega^2-\re{\omega_0}^2}{4\re{\omega_0}\im{\omega_0}}\Biggl(1+\frac{(\re{\omega_0}-\re{\omega_1})^2}{32\im{\omega_0}^2}
\\\nonumber&&\qquad\qquad-\frac{3\im{\omega_0}-\im{\omega_1}}{24\im{\omega_0}}\Biggr)
-\frac{\re{\omega_0}-\re{\omega_1}}{16\im{\omega_0}}
\\\nonumber&& -\frac{(\omega^2-\re{\omega_0}^2)^2}{16\re{\omega_0}^3\im{\omega_0}}\left(1+\frac{\re{\omega_0}(\re{\omega_0}-\re{\omega_1})}{4\im{\omega_0}^2}\right)
\\\nonumber&& +\frac{(\omega^2-\re{\omega_0}^2)^3}{32\re{\omega_0}^5\im{\omega_0}}\Biggl(1+\frac{\re{\omega_0}(\re{\omega_0}-\re{\omega_1})}{4\im{\omega_0}^2}
\\\nonumber&&\qquad +\re{\omega_0}^2\Biggl(\frac{(\re{\omega_0}-\re{\omega_1})^2}{16\im{\omega_0}^4}
\\&&\qquad\qquad -\frac{3\im{\omega_0}-\im{\omega_1}}{12\im{\omega_0}}\Biggr)\Biggr)+ \Order{\frac{1}{\ell^3}}.
\label{eq:gbsecondorder}
\end{eqnarray}
Here, $\omega_0$ represents the dominant quasinormal mode, while $\omega_1$ corresponds to the first overtone.

It is crucial to recognize that the WKB method, despite its widespread applicability, may fail under certain conditions, even for high multipole numbers where it is generally expected to perform well. This limitation arises when the effective potential deviates from the standard centrifugal form, $f(r) \ell (\ell + 1)/r^2$, particularly in scenarios involving higher-curvature corrections. In such cases, perturbative instabilities may emerge (see \cite{Konoplya:2020bxa,Takahashi:2011du,Gleiser:2005ra,Takahashi:2011qda,Konoplya:2017lhs,Takahashi:2010gz,Konoplya:2017ymp,Dotti:2005sq}).

Instances where the WKB method encounters difficulties in the eikonal limit or produces incomplete results are systematically reviewed in \cite{Konoplya:2022gjp,Bolokhov:2023dxq,Konoplya:2020bxa}. Notably, the WKB approach fails to capture the second branch of quasinormal modes that emerge in the spectra of asymptotically de Sitter black holes \cite{Konoplya:2022gjp}.
These modes represent the spectrum of the empty de Sitter spacetime \cite{Lopez-Ortega:2012xvr,Lopez-Ortega:2007vlo}. Therefore, the conclusions drawn in this analysis pertain exclusively to the Schwarzschild branch of modes, which can be reliably determined through the WKB method.

Under certain conditions, rotating black holes or charged fields in the vicinity of a charged black hole can exhibit the phenomenon of superradiance \cite{1971JETPL..14..180Z,Starobinskil:1974nkd,Starobinsky:1973aij,Bekenstein:1998nt,East:2017ovw,Hod:2012zza,Konoplya:2008hj,Bekenstein:1998nt,Konoplya:2014lha,Richarte:2021fbi,Zhu:2014sya}. This remarkable effect manifests when the reflection coefficient exceeds unity, implying that energy is extracted from the black hole during the scattering process. Superradiance occurs when the incoming radiation interacts with the rotating or charged black hole in such a way that the outgoing wave carries away more energy than the incident one.  

The emergence of superradiance introduces a significant complication, as it lies beyond the domain of applicability of the WKB approximation. The core assumption underpinning the WKB method presumes that the reflection coefficient remains bounded by unity, ensuring that the wave amplitude does not experience anomalous amplification during its evolution. Consequently, the correspondence between quasinormal modes and grey-body factors, which typically holds in the standard scattering scenario, ceases to be valid in the presence of superradiance.  

In such regimes, alternative approaches or numerical methods must be employed to accurately capture the behavior of perturbations and evaluate the scattering characteristics of the system. This deviation highlights the intricate and nontrivial dynamics governing energy extraction processes from black holes, an area that continues to be the subject of extensive theoretical investigation and exploration.

\section{Schwarzschild-de Sitter case}

Quasinormal modes of a massive scalar field in asymptotically de Sitter spacetimes can be determined with high precision using the higher-order WKB method. This method has demonstrated excellent agreement with alternative approaches, as highlighted in \cite{Fontana:2020syy,Dubinsky:2024hmn,Aragon:2020tvq}. As a result, quasinormal modes can either be extracted from previous works where they were computed with high accuracy or, with negligible loss of precision, derived using the higher-order WKB approximation enhanced by Padé approximants.  

In this work, we present plots of the grey-body factors calculated via the higher-order WKB method, alongside those obtained through the established correspondence with quasinormal modes. Figs. \ref{fig:Lambda005L0}-\ref{fig:Lambda005L2} reveal that as the mass of the scalar field increases, the discrepancy between the two approaches diminishes. Interestingly, as the mass parameter $\mu$ tends toward infinity, the error in the correspondence appears to converge to a minute value, which, while exceptionally small, does not necessarily vanish entirely.  

Conversely, the discrepancy in the correspondence grows with increasing multipole number $\ell$. This behavior is expected, as the influence of the field's mass diminishes relative to the dominant contribution of angular momentum at high $\ell$. In essence, the accuracy of the correspondence is governed not solely by the magnitude of the scalar field’s mass but by the interplay between the mass and the multipole number. Specifically, when
\begin{equation}
\mu M \gg \ell,
\end{equation}
one can confidently assert that the WKB method provides reliable results, and the correspondence reproduces grey-body factors with remarkable precision. This accuracy significantly surpasses that achieved for massless fields, as demonstrated in \cite{Bolokhov:2024otn,Skvortsova:2024msa,Konoplya:2024vuj}, where introducing a non-zero mass reduces the error by several orders of magnitude.  

Moreover, as depicted in Fig. \ref{fig:mu1point5}, the presence of a cosmological constant enhances the accuracy of the correspondence. However, in the near-extremal regime, this improvement is not strictly monotonic. In cases where the cosmological constant is small, and the scalar field mass is likewise minimal, the precision of the correspondence approaches that observed for massless fields in asymptotically flat black hole backgrounds.  

Ultimately, this study underscores that the accuracy of the WKB method and its correspondence with quasinormal modes depends sensitively on both the mass of the scalar field and the cosmological constant. These factors collectively contribute to reducing errors and ensuring the reliable computation of grey-body factors across a broad parameter space.

\section{Other asymptotically de Sitter black holes}

A similar phenomenon is observed for other asymptotically de Sitter black holes. In this work, we illustrate this behavior using the Reissner-Nordström-de Sitter (RNdS) black hole as a representative example. The metric function for the RNdS black hole is given by:
\begin{equation}\label{RNdSsolution}
f(r) = 1 - \frac{2 M}{r} + \frac{Q^2}{r^2} - \frac{\Lambda r^2}{3},
\end{equation}
where $M$ denotes the mass of the black hole, $Q$ represents its electric charge, and $\Lambda$ is the cosmological constant.  

As illustrated in Fig. \ref{fig:mu1point5Charged}, the effect of the electric charge $Q$ mirrors that of the cosmological constant $\Lambda$. Specifically, increasing the black hole charge reduces the relative error in the correspondence. This trend suggests that, akin to the role of $\Lambda$, the charge effectively enhances the accuracy of the correspondence, leading to a notable improvement in precision. For sufficiently large scalar field masses, the accuracy of this correspondence improves by several orders of magnitude across a variety of asymptotically de Sitter metrics, including the Schwarzschild-de Sitter and Reissner-Nordström-de Sitter spacetimes. 

A similar situation arises in the context of black holes that are not strictly asymptotically de Sitter but exhibit asymptotically de Sitter-like behavior. One notable example of this occurs in the Mannheim-Kazanas solution \cite{Mannheim:1988dj}, where the metric function takes the form:  
\begin{equation}\label{RNdSsolution}
f(r) = 1 - \frac{2 M}{r} - k r.
\end{equation}  
In this expression, the parameter $k > 0$ introduces a de Sitter-like horizon at some finite distance from the black hole. This horizon reflects the asymptotic expansion associated with the Mannheim-Kazanas spacetime, drawing parallels with cosmological horizons found in pure de Sitter space.  

For this configuration, the quasinormal modes (QNMs) can be determined analytically in the limit of large $\mu$, where $\mu$ represents the mass of the perturbing field. The analytical approach is facilitated by the simplicity of the effective potential near the peak and the application of WKB techniques.  

The radial location of the maximum of the effective potential is given by:  
\begin{equation}
r_{max} = -\frac{2 M}{\sigma }-\frac{(2 \sigma +1) \left(\ell ^2+\ell
   -\sigma \right)}{2 M \mu ^2}+O\left(\frac{1}{\mu
   }\right)^4,
\end{equation}  
where the dimensionless parameter $\sigma$ is defined by $\sigma^2 = 2 k M$. 

By expanding the potential in inverse powers of the field mass $\mu$ and applying the WKB approximation, the quasinormal spectrum can be derived to higher orders. The resultant expression for the frequency has the form
\begin{widetext}
\begin{equation}
\begin{aligned}
\omega_{n} = & \ \mu  \sqrt{2 \sigma +1}+\frac{\left(n+\frac{1}{2}\right)
   \sqrt{\sigma ^3 (2 \sigma +1)}}{2 M} \\
& -\frac{\sigma ^2
   \sqrt{2 \sigma +1} \left(12 \left(n+\frac{1}{2}\right)^2
   (2 \sigma +1)-32 \ell ^2-32 \ell -18 \sigma
   -5\right)}{256 M^2 \mu } \\
& +\frac{\left(n+\frac{1}{2}\right)
   \sigma ^{5/2} (2 \sigma +1)^{3/2} \left(92
   \left(n+\frac{1}{2}\right)^2 (2 \sigma +1)-662 \sigma
   +5\right)}{8192 M^3 \mu ^2}+O\left(\frac{1}{\mu
   }\right)^3.
\end{aligned}
\end{equation}

\end{widetext}

However, it is crucial to clarify the interpretation of overtone numbers in the context of the correspondence. When employing the fundamental ($n=0$) and first overtone ($n=1$) frequencies, this does not imply that we are considering the two longest-lived quasinormal modes. The overtone number $n$ pertains exclusively to the Schwarzschild-like branch of modes. This branch describes the resonances associated with the black hole's event horizon, whereas the de Sitter branch – representing modes governed by the cosmological horizon – is disregarded in this analysis.  

Thus, while the correspondence leverages the Schwarzschild branch for precise determination of grey-body factors, the de Sitter modes, which decay more rapidly, are intentionally excluded from the calculation. This selective focus ensures that the methodology remains consistent and retains high accuracy even as the underlying spacetime configuration varies with the inclusion of charge or cosmological terms.

\vspace{3mm}
\section{Conclusions}

In this work, we have extended the analysis of the correspondence between quasinormal modes and grey-body factors to asymptotically de Sitter black holes, focusing on perturbations of massive scalar fields. By employing the higher-order WKB method enhanced by Padé approximants, we demonstrated that the correspondence holds with exceptional accuracy when the mass of the field is sufficiently large.

Our results indicate that the relative error between grey-body factors derived directly from WKB calculations and those obtained through correspondence with QNMs decreases significantly with increasing scalar field mass. This trend persists across different multipole numbers, albeit with diminishing accuracy at higher $\ell$, where angular momentum becomes more influential than the mass of the field.

Additionally, we observed that the presence of a cosmological constant enhances the precision of the correspondence, although this improvement is not strictly monotonic in near-extremal regimes. The analysis was further extended to Reissner-Nordström-de Sitter and Mannheim-Kazanas black holes. The introduction of electric charge similarly reduced the relative error in grey-body factors.

Importantly, we clarified that the overtone numbers referenced in the correspondence pertain exclusively to the Schwarzschild branch of modes, with the de Sitter branch being disregarded. This selective focus ensures that the correspondence accurately describes the scattering properties associated with the black hole’s event horizon, without contributions from modes dominated by the cosmological horizon.

Our findings contribute to a deeper understanding of the interplay between dynamical resonances and scattering phenomena in black hole spacetimes, providing a robust framework for analyzing massive field perturbations in cosmological settings. Future work may explore extending this analysis to more complex black hole configurations, such as those arising in higher-dimensional or modified gravity theories.

\acknowledgments{
The author acknowledges Roman Konoplya for useful and encouraging discussions.

\bibliography{bibliography}

\end{document}